\documentclass[fleqn,usenatbib]{rasti}

\usepackage{newtxtext,newtxmath}

\usepackage[T1]{fontenc}

\DeclareRobustCommand{\VAN}[3]{#2}
\let\VANthebibliography\thebibliography
\def\thebibliography{\DeclareRobustCommand{\VAN}[3]{##3}\VANthebibliography}

\usepackage{graphicx}	
\usepackage{amsmath}	
\usepackage{mathtools}
\usepackage{dsfont}

\renewcommand{\vec}[1]{\boldsymbol{#1}}

\title[Robustly measuring field level statistics]{Methods for robustly measuring the minimum spanning tree and other field level statistics from galaxy surveys}

\author[K. Naidoo and O. Lahav]{
Krishna Naidoo$^{1,2}$\thanks{E-mail: \href{mailto:krishna.naidoo@port.ac.uk}{krishna.naidoo@port.ac.uk}}
and Ofer Lahav$^{2}$
\\
$^{1}$Institute of Cosmology and Gravitation, University of Portsmouth, Burnaby Road, Portsmouth, PO1 3FX, UK\\
$^{2}$Department of Physics \& Astronomy, University College London, Gower Street, London WC1E 6BT, UK
}

\date{Accepted XXX. Received YYY; in original form ZZZ}

\pubyear{\the\year{}}

\begin{document}
\label{firstpage}
\pagerange{\pageref{firstpage}--\pageref{lastpage}}
\maketitle

\begin{abstract}
Field level statistics, such as the minimum spanning tree (MST), have been shown to be a promising tool for parameter inference in cosmology. However, applications to real galaxy surveys are challenging, due to the presence of small scale systematic effects and non-trivial survey selection functions. Since many field level statistics are `hard-wired', the common practice is to forward model survey systematic effects to synthetic galaxy catalogues. However, this can be computationally demanding and produces results that are a product of cosmology and systematic effects, making it difficult to directly compare results from different experiments. We introduce a method for inverting survey systematic effects through a Monte Carlo subsampling technique where galaxies are assigned probabilities based on their galaxy weight and survey selection functions. Small scale systematic effects are mitigated through the addition of a point-process smoothing technique called jittering. The inversion technique removes the requirement for a computational and labour intensive forward modelling pipeline for parameter inference. We demonstrate that jittering can mask small scale theoretical uncertainties and survey systematic effects like fibre collisions and we show that Monte Carlo subsampling can remove the effects of survey selection functions. We outline how to measure field level statistics from future surveys.
\end{abstract}




\section{Introduction}

The next generation of cosmological galaxy surveys; such as the Dark Energy Spectroscopic Instrument\footnote{\href{https://www.desi.lbl.gov/}{https://www.desi.lbl.gov/}} (DESI; \citealt{DESI2016}), \emph{Euclid}\footnote{\href{https://www.euclid-ec.org/}{https://www.euclid-ec.org/}}\citep{LaureijsEuclid2011}, the Rubin Observatory's Legacy Survey of Space and Time\footnote{\href{https://www.lsst.org/}{https://www.lsst.org/}} (LSST; \citealt{IvezicLSST2019}) and the Wide field Spectroscopic Instrument\footnote{\href{https://www.wstelescope.com/}{https://www.wstelescope.com/}} (WST; \citealt{Mainieri2024}); will provide a deeper and more resolved view of the universe than has even been available in the past. To maximise the information extracted from this data requires exploring statistics beyond the two-point correlation function (2PCF), including the 2PCF in different densities or cosmic web environments \citep{Bonnaire2022, Bonnaire2023, Paillas2023, Massara2023}, and higher-order statistics; such as the minimum spanning tree \citep[MST;][]{NaidooBeyond2020, NaidooQuijote2022}, Minkowski functionals \citep{Liu2022, Liu2023}, critical points \citep{Moon2023}, topological persistence homology \citep{Pranav2017, Jalali2023}, graph representations \citep{Makinen2022}, voids \citep{Kreisch2022}, wavelet scattering transforms \citep{Valogiannis2022}, and methods directly using field level inference \citep{Fluri2018, Lemos2023, Jeffrey2024}. However, many of these techniques cannot be modelled analytically, and instead require predictions from simulation suites exploring a large cosmological parameter space. To complicate the matter, since these are often performed on galaxy catalogues, additional modelling are required to explore the halo-to-galaxy relation, often carried out by marginalising over halo occupation distribution (HOD) parameters \citep{Zheng2005}.

One such example of a field level statistic is the MST, a highly optimised graph built from a set of points and first introduced to astronomy by \citet{Barrow1985}. In graph theory, a tree is defined as a loop-free structure, while `spanning' refers to a graph connecting all points in a single structure. The MST is the spanning tree with the shortest possible total length. This optimisation leads to a graph, the MST, that very effectively traces filaments in the cosmic web \citep[see][]{Libeskind2018} and, more recently, has been shown to be highly sensitive to neutrino mass \citep{NaidooQuijote2022}. Unlike the $N$-point correlation function, the MST is sensitive to density, an effect that cannot be removed by the inclusion of randoms to account for survey selection effects. Additionally, because the MST operates directly on galaxies, we cannot exclude small scales by pixelising the galaxy distribution onto a grid. Furthermore, there is no mechanism to directly incorporate galaxy weights, which are typically used to correct for observational systematic effects, into the MST graph construction. Exploiting `hard-wired' statistics -- i.e. sophisticated algorithms like the MST or machine learning with fixed configuration -- presents a unique challenge for observational cosmology. While the MST is the statistic of interest in this paper, hard-wired artificial intelligence and machine learning algorithms, such as graph neural networks, will benefit from resolving these current limitations.

The standard approach for using hard-wired statistics like the MST for parameter inference would be to forward model survey systematic effects and selection functions to synthetic galaxy catalogues. However, forward modelling survey systematic effects to the levels of accuracy needed for the MST has yet to be performed and would be computationally and labour intensive,. The benefits of forward modelling is that it enables inference in scenarios that appear intractable, see \citet{Lemos2023} and \citet{Jeffrey2024} for powerful demonstrations of field level simulation based inference techniques used in cosmology. However, in forward modeling, we lose the ability to clearly delineate a cosmological measurement from survey systematic effects -- we can only compare and contrast measurements at the parameter level. Even if different aspect of the forward model are turned on/off, we will only be able to compare this at the parameter level and therefore cannot directly compare survey measurements to simulations. Being able to look and directly interpret measurements will become particularly important when measurements are in `tension' with other probes or if they suggest the discovery of new physics.

In this paper, we outline how to make robust measurements of the MST and other hard-wired field level statistics from real galaxy surveys. We illustrate how to marginalise over survey selection functions, such as the redshift selection function and variabilities in completeness, and how to mitigate small scale systematic effects. The techniques introduced in this paper remove the necessity to forward model survey geometry, selection functions and systematic effects to synthetic galaxy catalogues. Our techniques involve inverting survey systematic effects on real data and comparing to synthetic galaxy catalogues without survey systematic effects. Since the methods are imposed at the catalogue level they can be applied generally to any technique. In section~\ref{sec_methods} we describe the mock galaxy catalogues used, the summary statistics measured and the methods for inverting survey selection functions and mitigating small scale systematic effects. In section~\ref{sec_results} we validate the techniques, in~\ref{sec_discussion} we discuss the methods and outline how to use them for measuring the MST from future galaxy redshift surveys and in~\ref{sec_summary} we summarise the results of the paper.

\section{Methods}
\label{sec_methods}

In this section we summarise the simulations, summary statistics and techniques used throughout this paper. We then introduce the Monte Carlo subsampling technique used to indirectly apply galaxy weights and mitigating survey systematic effects at the catalogue level.

\subsection{Simulations}

\subsubsection{Millennium XXL galaxy lightcone}
\label{sec_galcat}

We use the Millennium XXL galaxy lightcone catalogue of \citet{Smith2022}. The lightcones are produced from the Millennium XXL simulation, computed in a box of $3\,h^{-1}{\rm Gpc}$ in a flat $\Lambda$CDM cosmology \citep{Planck2020} with matter density $\Omega_{\rm m}=0.25$, dark energy density $\Omega_{\rm \Lambda}=0.75$, amplitude of fluctuations at $8\,h^{-1}{\rm Mpc}$ $\sigma_{8}=0.9$, the Hubble constant $H_{0}=73$ and primordial spectral tilt $n_{\rm s}=1$. For this analysis we limit the catalogue to halos of mass greater than $\geq 10^{13} h^{-1}M_{\odot}$ and only consider galaxies either in a single octant (i.e. $0^{\circ} \le {\rm RA} \le 90^{\circ}$ and $0^{\circ}\le {\rm Dec.}\le 90^{\circ}$) or inside the Baryonic Oscillation Spectroscopic Survey (BOSS) LOWZ north footprint. For the octant there are approximately $180,000$ synthetic galaxies in the mock catalogue, while for the BOSS LOWZ north footprint there are approximately $210,000$ synthetic galaxies.

\subsubsection{L\'{e}vy flight random walk distribution}

We use a set of L\'{e}vy flight random walk simulations. Unlike cosmological simulations the points produced have no higher-order information, since the only input is the size of the steps used in the random walk, and they are fast to generate -- making them a very useful tool for testing the effects of small scale differences on the MST. We use two sets of distributions, a standard technique which we will refer to as L\'{e}vy flight \citep[LF;][]{Mandelbrot1982} and the Adjusted L\'{e}vy flight (ALF) which we developed in \citet{NaidooBeyond2020}. These distributions are approximately equal on large scales but have significantly different small scale distributions. Both the LF and ALF random walk distributions are implemented in the \texttt{MiSTree} python package \citep{NaidooMistree2019}. The parameters used for the LF are the minimum step-size $t_{0}=0.2$ and tilt $\alpha=1$, and for the ALF are the step-size parameters $t_{0}=0.325$ and $t_{s}=0.015$, and the step-size shape parameters $\alpha=1.5$, $\beta=0.45$ and $\gamma=1.3$. Both distributions are produced in a periodic box of length $75\,h^{-1}{\rm Mpc}$.
 
\subsection{Summary statistics}

We describe the summary statistics used in this paper and the jackknife resampling technique used for error estimation.

\subsubsection{Two-point correlation function}

The 2PCF \citep{Peebles1980}, is a tried-and-tested method for computing clustering properties of an input dataset. It is widely used in cosmology and generally at the forefront of any observation in large scale structure. Analytical predictions for observations can be made from a given power spectra and galaxy bias prescription. Galaxy survey geometries and systematic effects are routinely mitigated through the use of randoms to account for the anisotropic and inhomogeneous distribution of galaxies from a galaxy survey, while small scale systematic effects can be masked by limited the analysis to separations beyond some input scale (say $r_{\min}$). The 2PCF can be computed from the \citet{LandySzalay1993} estimator
\begin{equation}
    \xi(r) = \left(\frac{n_{R}}{n_{D}}\right)^{2}\frac{DD(r)}{RR(r)} - 2\left(\frac{n_{R}}{n_{D}}\right)\frac{DR(r)}{RR(r)} + 1,
\end{equation}
where $r$ is the distance between points, $DD(r)$ the number of galaxy-galaxy pairs at a distance $r$, $DR(r)$ the number of galaxy-random pairs, $RR(r)$ the number of random-random pairs, $n_{D}$ the mean number density of galaxies and $n_{R}$ the mean number density of randoms.

To compute the 2PCF multipoles we measure the 2PCF binned according to the distance parallel to the line-of-sight of the observer $s_{\parallel}$ and the distance perpendicular to the line-of-sight $s_{\perp}$
\begin{equation}
    \xi(s_{\parallel}, s_{\perp}) = \left(\frac{n_{R}}{n_{D}}\right)^{2}\frac{DD(s_{\parallel}, s_{\perp})}{RR(s_{\parallel}, s_{\perp})} - 2\left(\frac{n_{R}}{n_{D}}\right)\frac{DR(s_{\parallel}, s_{\perp})}{RR(s_{\parallel}, s_{\perp})} + 1.
\end{equation}
The multipoles are then computed from
\begin{equation}
    \xi_{\ell}(s)=\frac{2\ell+1}{2}\int^{\pi}_{0}\sqrt{1-\mu^{2}}\, \xi(s_{\parallel}, s_{\perp})\,P_{\ell}(\mu){\rm d}\theta,
\end{equation}
where $s=\sqrt{s_{\parallel}^{2}+s_{\perp}^{2}}$, $\mu=\cos\theta$ and $P_{\ell}$ is the Legendre polynomial. The monopole is computed by setting $\ell=0$, the quadrupole with $\ell=2$ and hexadecapole with $\ell=4$.

\subsubsection{Minimum spanning tree}

The MST is computed on a 3-dimensional distribution of points using the \texttt{MiSTree} python package \citep{NaidooMistree2019}. From the constructed MST graph we measure:
\begin{itemize}
    \item Degree ($d$): the number of connected edges to each node/point.
    \item Edge length ($l$): the length of each edge.
    \item Branches: edges connected in chains with intermediate nodes of $d=2$. From branches we measure:
    \begin{itemize}
        \item Branch length ($b$): the total length of member edges.
        \item Branch shape ($s$): the straight line distance between the branch ends divided by the total branch length.
    \end{itemize}
\end{itemize}
We are interested in the probability distribution function (PDF) of $d$, $l$, $b$ and $s$ that are obtained by binning the MST statistics into histograms. While this is not an exhaustive set of statistics to measure from the MST, we have found the PDFs to be highly constraining \citep{NaidooQuijote2022} for cosmology, in particular the PDFs of edge length $l$ and branch length $b$.

\subsubsection{Error estimation with jackknife resampling}

We use the jackknife resampling technique to compute errors. We first divide our original dataset into $N_{\rm JK}$ jackknife regions. For datasets inside a periodic box, the box is divided into $N_{\rm JK}^{1/3}$ segments along each axis, creating $N_{\rm JK}$ smaller cubic boxes within the full box. For datasets from a lightcone we segment the footprint into $N_{\rm JK}$ regions using the binary-partition method in Naidoo et al.~(in prep). We then compute the statistics of interest with points in the $i^{\rm th}$ jackknife segment removed. This gives us $N_{\rm JK}$ estimates of the statistic, which we will refer to as $\vec{y}$, where we take the mean 
\begin{equation}
    \bar{\vec{y}} = \frac{1}{N_{\rm JK}}\sum_{i=1}^{N_{\rm JK}} \vec{y}_{i},
\end{equation}
to be the statistic for the full sample. This can be validated, to test for biases from removing a jackknife segment, by comparing to the statistic measured on the full sample. The jackknife estimate of variance for the statistics is given by
\begin{equation}
    \vec{\Delta y}^{2}_{\rm JK} = \frac{N_{\rm JK}-1}{N_{\rm JK}}\sum_{i=1}^{N_{\rm JK}}(\vec{y}_{i}-\bar{\vec{y}})^{2}.
\end{equation}
Note, in practice we can compute the variance by assuming the samples are independent and then correcting the variance obtained by multiplying by the Jackknife pre-factor $(N_{\rm JK}-1)^{2} / N_{\rm JK}$.

\subsection{Survey systematic effects and theoretical limitations}

Cosmological datasets, whether they come from real galaxy surveys, or theoretical data from simulations will always come with uncertainties. It is important that these uncertainties are accounted for, particularly within an inference pipeline. Accounting for these uncertainties is particularly challenging for hard-wired algorithms considered in this paper. In this section, we explore how galaxy surveys and theoretical data from $N$-body simulations can introduce uncertainties in both our model and observational measurements.

\subsubsection{Small scale theoretical unknowns}

Simulations will be integral to making predictions for higher-order cosmic web statistics, but simulations have uncertainties of their own that need to be accounted for and mitigated in any cosmological analysis. The most problematic are uncertainties on small scales, that are dictated by the force and mass resolution of the simulation -- an effect that can be larger for approximate $N$-body solvers such as COLA \citep{Tassev2013}. This can result in simulations where only massive halos are resolved reliably, making predictions difficult for observations of galaxies in small mass halos. This is conceptually a rather simple limitation to solve -- simply discard low mass galaxies or use higher resolution simulations. A more pressing issue is the uncertainty in the physics of small scales, $\lesssim10h^{-1}{\rm Mpc}$. This is the domain where we expect baryonic physics and feedback mechanisms to be important and where there remains uncertainties on the galaxy formation prescriptions used for hydrodynamic simulations. How these scales are incorporated in any cosmological analysis needs to be treated with great care. Uncertainty on small scales are further enhanced by the fact that simulations with hydrodynamic galaxy formation physics are expensive to run, and not always available, especially for the large $N$-body simulation suites used for cosmological inference. Typically in cosmology, galaxies are `painted' onto dark matter only simulations using a halo-to-galaxy prescription; this is often constructed using HODs or semi-analytic halo abundance matching \citep[SHAM;][]{Vale2004} or both. Marginalising over the uncertainties in these semi-analytic models will remove some of the uncertainties but eventually these models will break down. Therefore, we can only ever expect simulations to provide reliable galaxy catalogues up to some theoretical lower bound, a lower bound which cannot (at the moment) be masked from a hard-wired algorithm.

\subsubsection{Small scale observational systematic effects}

In observations, small scales provide a different challenge. In spectroscopic surveys, galaxies that are close together on the sky may not be assigned individual fibres for their respective spectras to be measured. This problem, known as fibre collisions, leads to galaxies that cannot be assigned correct redshifts. To correct for these missing galaxies, the nearest galaxy is often re-weighted, to account for the fact that there are more galaxies nearby. Nevertheless, this means on angular scales smaller than the fibre collision length, the distribution of galaxies cannot be trusted. From photometric surveys, a different problem occurs on small scales, and that is a problem caused by the blending and crowding of galaxies in close proximity. In effect, galaxies become difficult to resolve, making photometry and shape measurements difficult and unreliable. This can lead to inaccuracies in photometric redshift estimates and can introduce systematic errors in shape measurements relevant for weak lensing studies. In both cases we cannot trust the information on these small scales and need to be able to remove them.

\subsubsection{Radial selection function}

Depending on the characteristics of the survey, galaxies will be observed with a characteristic radial redshift selection function. This is substantially different from what we have from simulations where the distribution of galaxies is homogeneous. Not only is the distribution inhomogeneous, but so too are the galaxy properties. For example we will often find only small galaxies being observed at lower redshifts and only massive galaxies at higher redshifts. If a hard-wired algorithm is applied directly to this distribution of galaxies, it may be overwhelmed by the inhomogeneous redshift selection function and the biased galaxy redshift evolution. While the latter can be addressed by modeling the galaxy-halo connection, the former diminishes the relevant cosmological information in the hard-wired algorithm, with a significant portion merely capturing survey systematics. This complicates comparisons across different experiments, making it difficult to disentangle true cosmological signals from survey-induced effects.

\subsubsection{Angular selection function}

Galaxy surveys will not be observed isotropically, but will instead be observed along the galaxy survey's footprint. Surveys typically have non-trivial boundaries which may induce significant biases in the results from hard-wired algorithms. In addition, galaxies observed across the sky will be subject to the seeing conditions during observations and galactic foregrounds, such as zodiacal light and stellar density from our own galaxy that will induce anisotropies in the distribution of galaxies observed. This effect will need to be replicated in a forward modelled approach.

\subsection{Monte Carlo subsampling}
\label{sec_monte_carlo_prob}

In $2$-point (or even $N$-point) statistics galaxy survey systematic effects are mitigated by assigning weights to each galaxy. The weights capture observational biases, such as the angular and radial selection functions, and on small scales can account for missing galaxies, for e.g. from fibre collisions or blending of galaxies in crowded fields. Although obtaining galaxy weights from cosmological surveys is a laborious task, incorporating these weights into $2$-point clustering measurements is straightforward. However, for the MST and hard-wired algorithms, it is unclear how weights should be included. The only approach to apply these promising techniques to real data for parameter inference is to carry out a full forward modelling approach. This will mean the creation of realistic galaxy mock catalogues for a given survey and accurate forward modelling of the survey's systematic effects and selection functions. This is both computationally expensive and labour intensive, and because the measurements are a product of both cosmological and systematic information, it becomes difficult to interpret results and thereby the parameter constraints they leads to. For these reasons, we look for a different approach, one where survey systematic effects are mitigated at the catalogue level -- removing the need for the direct inclusion of weights in the calculation of the MST or other hard-wired algorithms.

\subsubsection{Systematic weighted probabilities}

To incorporate systematic effects at the catalogue level we subsample a galaxy catalogue, treating the galaxy weights $w_{\rm g}$ (which are constructed to include various systematic properties of a galaxy survey) as probabilities $\mathds{P}_{\rm g}=w_{\rm g}$. To subsample the galaxies we draw a random uniform number $u\in [0,1]$ and only keep galaxies when $\mathds{P}_{\rm g}\le u$. We then measure the statistics of interest, which we will refer to as $\vec{d}$ (a vector of length $n_{\rm d}$), measured for the subsampled galaxy distribution, which we will refer to as $\tilde{\vec{d}}$. We repeat the process over many iterations, taking the mean of the measured statistic ($\vec{d}_{n}$) evaluated over the ensemble of $n$ iterations,
\begin{equation}
    \vec{d}_{n} = \frac{1}{n} \sum_{i=1}^{n} \tilde{\vec{d}}_{i},\quad{\rm where}\quad \lim_{n\to \infty} \vec{d}_{n} \equiv \vec{d},
\end{equation}
which is only true if $\tilde{\vec{d}}$ is unbiased. In this way we are able to incorporate the galaxy weights indirectly, since when $n\to\infty$ a galaxy will be sampled $\approx n\mathds{P}_{\rm g}$, averaged over $n$ iterations gives the galaxy the correct weight $\approx \mathds{P}_{\rm g} = w_{\rm g}$. This process is akin to the superposition of different quantum states in quantum mechanics, each iteration gives us a discrete set of observations following an underlying probability distribution function. A distribution that becomes clearer once the average is taken over many discrete observations.

Treating $w_{\rm g}$ as probabilities $\mathds{P}_{\rm g}$ only make sense if $w_{\rm g} \in [0, 1]$, if $w_{\rm g}>1$ then this galaxy will always be assigned the incorrect weight, biasing the analysis. To avoid this potential issue, we replicate galaxies with a weight greater than one by $n_{g}$ times, where $n_{g} = \lceil w_{\rm g}\rceil$, and $\lceil x\rceil$ is the ceiling function (which rounds $x$ to the nearest integer greater than $x$). The new weights for the replicated galaxies are now $\mathds{P}_{\rm g}=w_{\rm g}/n_{\rm g}$. This results in subsampled galaxies that have the correct weights. Galaxy weights greater than one, often occur when nearby galaxies cannot be observed, so it is important in practice that the replication of galaxies with weights greater than one happens in combination with the addition of a jitter (described below). This also ensures galaxy replicates do not lie on top of each other.

\subsubsection{Point-process smoothing with jittering}

To remove systematic effects and theoretical uncertainties at small scales, we add noise to the three dimensional position of the galaxy $\vec{p}$, a process we call jitter
\begin{equation}
    \vec{p}_{\rm J} = \vec{p} + \mathcal{N}(0,\,\sigma_{\rm J}\cdot\vec{I}_{3}).
\end{equation}
$\mathcal{N}$ is the normal distribution with mean zero and standard deviation given by the dot product of the three dimensional identity matrix $\vec{I}_{3}$ and jitter dispersion $\sigma_{\rm J}$. This process is the point-process equivalent to smoothing a field. Like smoothing, we expect this procedure to remove systematic effects at small scales and remove any sensitivity to small scales. This is crucial for the MST where small scales cannot be post-processed. Therefore, we must be able to hide galaxy survey systematic effects and theoretical uncertainties from simulations to be able to use the MST for parameter inference from galaxy surveys.

\subsubsection{Removing selection functions with inverse density weights}

Galaxy surveys have radial and angular selection functions, the former being characterised by the redshift galaxy density distribution $\rho(z)$ and the latter characterised by the surveys footprint/mask $M({\rm RA}, {\rm Dec})$ and completeness $C({\rm RA}, {\rm Dec})$. Both properties create variations in the distribution of galaxies, variations which are properties of the survey and are not cosmological in origin. In $2$-point or $N$-point statistics, the effect of the surveys selection function can be removed with the inclusion of randoms which allow one to measure the extra clustering with respect to a uniform sample. If we perform the MST, or other statistics, directly on galaxies we will get results which are a combination of both the cosmology and the survey selection function. This makes the analysis of such results difficult and interpretation challenging. To remove this issue we subsample galaxies to produce a uniform distribution of galaxies both radially and across the sky. This process involves applying what we call inverse density weights, which assigns a galaxy a probability that is a function of its radial and angular densities,
\begin{equation}
\mathds{P}_{\rm g} = w_{\rm g}\,\left(\frac{\rho_{\rm target}}{\rho(z_{\rm g})}\right)\,\left(\frac{C_{\rm target}}{C({\rm RA}_{\rm g}, {\rm Dec}_{\rm g})}\right).
\label{eq_probgal}
\end{equation}
Where the subscript $\rm g$ simply refers to the respective values for a single galaxy, $\rho_{\rm target}$ and $C_{\rm target}$ are target radial densities and target completeness values for the subsampled uniform distribution of galaxies. The galaxies being considered for subsampling should be choosen such that $\rho(z_{\rm g})\ge\rho_{\rm target}$ and $C({\rm RA}_{\rm g}, {\rm Dec}_{\rm g})\ge C_{\rm target}$. The effective density of the subsampled distribution is $\rho_{\rm eff} = \rho_{\rm target} C_{\rm target}$. This means theoretical predictions from simulations can aim to measure the MST or other hard-wired algorithm for a well defined galaxy distribution at an effective density $\rho_{\rm eff}$, a much simpler task than attempting to forward model the surveys complicated radial and angular selection functions.

\subsubsection{Convergence criteria}
\label{sec_convergence_criteria}

To estimate the statistics $\vec{d}_{n}$ requires applying the iterative probabilistic subsampling scheme over $n$ iterations. If the application of weights is unbiased then we expect for $n\to \infty$, $\vec{d}_{n}\equiv \vec{d}$. However, for this procedure to be tractable $n$ should be small but large enough for $\vec{d}_{n}\approx \vec{d}$. So, how do we determine the value of $n$? To understand this it is useful to first approach this assuming we have access to $\vec{d}$. One way to define $n$ is to look at the fractional difference 
\begin{equation}
    \epsilon_{f} = \frac{1}{n_{\rm d}}\sum_{j=1}^{n_{\rm d}}\frac{\vec{d}^{(j)}_{n}-\vec{d}^{(j)}}{\vec{d}^{(j)}},
\end{equation}
where the subscript $(j)$ refers to single elements across the data vector. We take $n$ to be when this fractional difference reaches some predefined threshold $\epsilon_{\rm thresh}\le\epsilon_{f}$. In reality we do not have access to $\vec{d}$, but we do have access to $\vec{d}_{n+1}$. We can then try to redefine the convergence by the iterative fractional difference
\begin{equation}
    \epsilon_{I} = \frac{1}{n_{\rm d}}\sum_{j=1}^{n_{\rm d}}\frac{\vec{d}^{(j)}_{n}-\vec{d}^{(j)}_{n+1}}{\vec{d}^{(j)}_{n+1}}.
\end{equation}
For $\epsilon_{f}$ the variance is fully described by the variance in $\vec{d}_{n}$ which in turn is described by the variance in $\tilde{\vec{d}}$,
\begin{equation}
    {\rm Var}\left(\vec{d}_{n}\right) = \frac{1}{n}{\rm Var}\left(\tilde{\vec{d}}\right).
\end{equation}
Therefore the variance of $\epsilon_{f}$ is given by
\begin{equation}
    {\rm Var}(\epsilon_{f}) = \frac{1}{n\,n_{\rm d}^{2}}\sum_{j=1}^{n_{\rm d}}\frac{{\rm Var}\Bigl(\tilde{\vec{d}}^{(j)}\Bigr)}{\Bigl(\vec{d}^{(j)}\Bigr)^{2}}.
    \label{eq_var_epf}
\end{equation}
The variance of $\epsilon_{I}$ is a little more complicated to compute since $\vec{d}_{n}$ and $\vec{d}_{n+1}$ are strongly correlated since the same $n$ samples are used to compute both of them. It is useful to remove this correlation and redefine $\epsilon_{I}$ as
\begin{equation}
    \epsilon_{I} = \frac{1}{n_{\rm d}(n+1)}\sum_{j=1}^{n_{\rm d}}\frac{\vec{d}_{n}-\tilde{\vec{d}}_{n+1}}{\vec{d}_{n+1}}.
\end{equation}
Assuming $\vec{d}_{n+1} = \vec{d} + \vec{\Delta d}_{n+1}$, we can Taylor expand the denominator to write it in terms of $\vec{d}^{(j)}$,
\begin{equation}
    \begin{split}
    \epsilon_{I} & = \frac{1}{n_{\rm d}\,(n+1)}\sum_{j=1}^{n_{\rm d}}\frac{\vec{d}_{n}^{(j)}-\tilde{\vec{d}}_{n+1}^{(j)}}{\vec{d}^{(j)} + \vec{\Delta d}_{n+1}^{(j)}},\\
    &\approx \frac{1}{n_{\rm d}\,(n+1)}\sum_{j=1}^{n_{\rm d}}\frac{\vec{d}_{n}^{(j)}-\tilde{\vec{d}}_{n+1}^{(j)}}{\vec{d}^{(j)}}\left(1 + \mathcal{O}\left(\frac{\vec{\Delta d}_{n+1}^{(j)}}{\vec{d}^{(j)}}\right)\right).
    \end{split}
\end{equation}
Finally, since $\vec{\Delta d}_{n+1} \ll \vec{d}$ we can make the approximation
\begin{equation}
    \epsilon_{I} \approx \frac{1}{n_{\rm d}(n+1)}\sum_{j=1}^{n_{\rm d}}\frac{\vec{d}_{n}^{(j)}-\tilde{\vec{d}}_{n+1}^{(j)}}{\vec{d}^{(j)}},
\end{equation}
and therefore the variance is given by
\begin{equation}
    {\rm Var}(\epsilon_{I}) \approx \frac{1}{n\,n^{2}_{\rm d}(n+1)}\sum_{j=1}^{n_{\rm d}}\frac{{{\rm Var}\left(\tilde{\vec{d}}^{(j)}\right)}}{\left(\vec{d}^{(j)}\right)^{2}}
\end{equation}
and therefore from eq.~\ref{eq_var_epf} we get
\begin{equation}
    {\rm Var}(\epsilon_{I}) \approx \frac{{\rm Var}(\epsilon_{f})}{n+1}.
\end{equation}
Since the expectation of $\epsilon_{f}$ and $\epsilon_{I}$ is completely described by their variance we can relate the two by their standard deviation and thus
\begin{equation}
    \epsilon_{f}\approx \sqrt{n+1}\,\epsilon_{I}.
\end{equation}
Which means we can define a threshold for $\epsilon_{f}$ without knowing $\vec{d}$.

\begin{figure*}
    \centering
    \includegraphics[width=\textwidth]{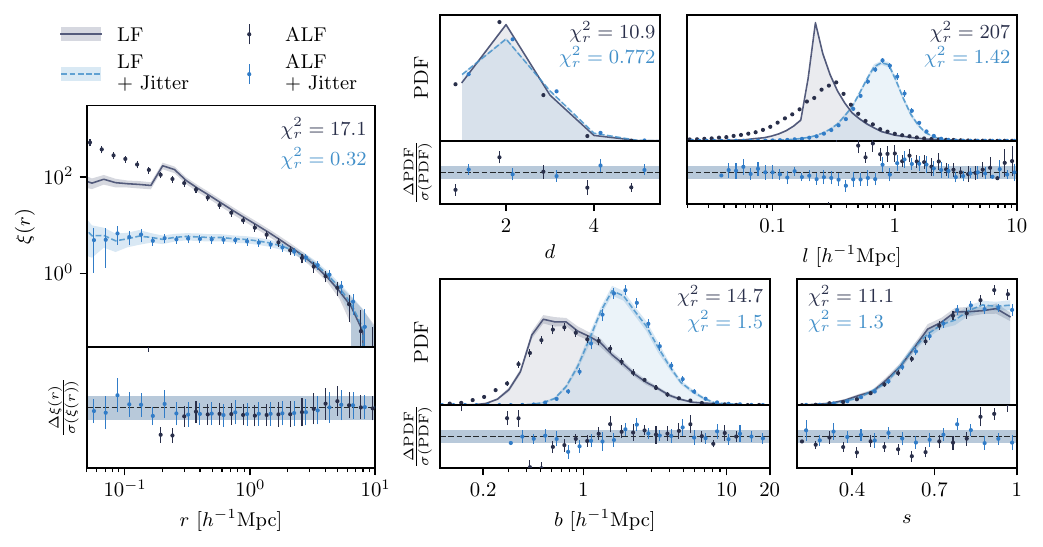}
    \caption{\label{fig_jitter_LF}The 2PCF (left subpanel) and MST distributions for $d$, $l$, $b$ and $s$ (right subpanels) for the random walk distributions LF and ALF are shown with and without a jitter (random Gaussian displacement). Without the jitter (dark blue), the LF and ALF show distinctively different distributions arising from differences in the 2PCF on small scales. For the MST the small scale differences percolate into the statistics and create signficant departures in the MST statistics on all scales. The discrepancies are characterised with a reduced $\chi^{2}$, which for the distributions without a jitter are high -- indicating significant discrepancies. With a jitter (light blue) the LF and ALF distributions are completely consistent in their 2PCF and MST statistics with $\chi^{2}_{r}\lesssim 1.5$. This demonstrates that adding a jitter can mask theoretical uncertainties on small scales.}
\end{figure*}

The choice of threshold here appears to be quite arbitrary, a more robust way to define this would be to define the significant difference
\begin{equation}
    \varepsilon_{f} = \frac{1}{n_{\rm d}}\sum_{j=1}^{n_{\rm d}}\frac{\vec{d}_{n}^{(j)}-\vec{d}^{(j)}}{\vec{\Delta d}^{(j)}},
\end{equation}
where $\vec{\Delta d}$ is the error associated with $\vec{d}$, which can be measured through jackknife resampling. Similarly, we will not know $\vec{\Delta d}$ a priori and will need to instead rely on an iterative estimator
\begin{equation}
    \varepsilon_{I} = \frac{1}{n_{\rm d}}\sum_{j=1}^{n_{\rm d}}\frac{\vec{d}^{(j)}_{n}-\vec{d}^{(j)}_{n+1}}{\vec{\Delta d}_{n+1}^{(j)}}.
\end{equation}
Applying the same logical steps presented before and taking the Taylor expansion of $1/\vec{\Delta d}_{n+1} \approx 1/\vec{\Delta d}$ we get the relation
\begin{equation}
    \varepsilon_{f}\approx \sqrt{n+1}\,\varepsilon_{I}.
\end{equation}

By calculating $\epsilon_{f}$ and $\varepsilon_{f}$ through the approximations from the iterative $\epsilon_{I}$ and $\varepsilon_{I}$ we can set thresholds on the absolute $\epsilon_{f}$ and $\varepsilon_{f}$ to estimate how many iterations $n$ are required to get the statistics to the desired level of precision. This can be carried out without knowing the true weighted $\vec{d}$ or its true error $\vec{\Delta d}$. Furthermore, this allows us to ensure uncertainties from the monte carlo subsampling iterations are small in comparison to the sample variance, i.e.~by ensuring $\varepsilon_{f}\ll 1$, which will allow us to maximise the statistical power of the data.

\section{Results}
\label{sec_results}

In this section we will explore the performance of the Monte Carlo subsampling and jittering procedure in removing survey systematic effects and masking theoretical uncertainties.

\subsection{Jittering performance}

\subsubsection{Masking small scale theoretical uncertainties}

We compare the 2PCF and MST for two random walk distributions LF \citep{Mandelbrot1982} and ALF \citep{NaidooBeyond2020}. The distributions have identical higher-order information since they are both random walks. On scales $\gtrsim 0.3\,h^{-1}{\rm Mpc}$ they, by design, have approximately equal 2PCF -- see the left subplot of Fig.~\ref{fig_jitter_LF}. Despite the agreement of the 2PCF at scales $r \gtrsim 0.3\,h^{-1}{\rm Mpc}$ the MST shows significant differences in the distribution of $d$, $l$, $b$ and $s$ (dark blue). This agreement between the LF and ALF distributions for the 2PCF and MST is indicated by the reduced $\chi^{2}$, where errors are computed via jackknife resampling and added in quadrature. The $\chi^{2}_{r}$ are extremely significant for the MST, especially for the edge lengths $l$ where $\chi^{2}_{r}=207$. The $\chi^{2}_{r}$ is also large for the 2PCF, at $\chi^{2}_{r}=17.1$, but this is due to the inclusion of small scales, if we limit the results to larger scales than $r \gtrsim 0.3\,h^{-1}{\rm Mpc}$ then it reduces to $\chi^{2}_{r}=0.35$. Unlike the 2PCF, we cannot simply ignore edges of the MST above some threshold scale as these scales percolate into other areas of the statistics, such as the branch shape and degree, where it is unclear how we could remove these scales retroactively. For this reason we introduce the jittering scheme, which adds a random `jitter' or displacement to the location of each point in the distribution. The jitter is defined with a jitter dispersion of $\sigma_{\rm J}=1\,h^{-1}{\rm Mpc}$. By adding a random jitter to the positions of the points, we are effectively applying a point-process equivalent of smoothing a field. The 2PCF with the jitter shows complete consistency between the LF and ALF distributions, with $\chi^{2}_{r}=0.32$ (light blue line for LF and points for ALF). Similarly we show the distributions of $d$ now has a $\chi^{2}_{r}=0.772$, $l$ a $\chi^{2}_{r}=1.42$, $b$ a $\chi^{2}_{r}=1.5$, and $s$ a $\chi^{2}_{r}=1.3$. The $\chi^{2}_{r}$ are now all consistent with the same distribution, showing that adding a jitter to the location of points can masks differences on small scales.

\begin{figure*}
    \centering
    \includegraphics[width=\textwidth]{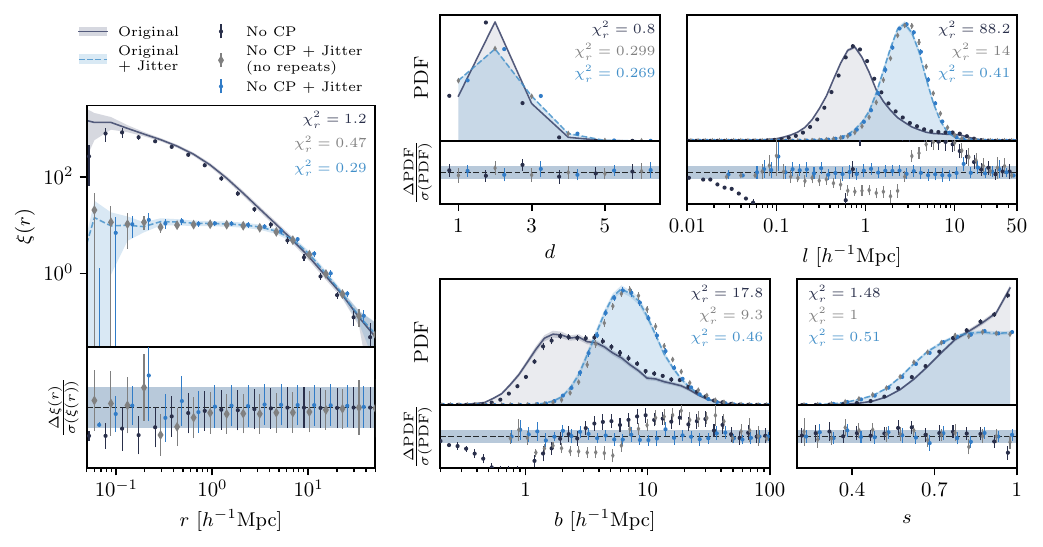}
    \caption{\label{fig_jitter_nocp}The 2PCF (left subpanel) and MST distributions for $d$, $l$, $b$ and $s$ (right subpanels) for galaxies from the original Millennium XXL ligthcone octant (lines) and without close pairs (No CP; an approximation for fibre collisions) are shown with and without a jitter (random Gaussian displacement). Without the jitter (dark blue), the original and No CP show consistent 2PCF but different MST distributions. The discrepancies for the MST are shown with measurements of the reduced $\chi^{2}$, which for the distributions without a jitter are high -- indicating significant discrepancies especially for the distribution of $l$ where $\chi^{2}_{r}=88.2$. With a jitter (light blue) the original and No CP distributions are completely consistent in their 2PCF and MST statistics where the $\chi^{2}_{r}\lesssim 0.5$. This only holds when the jitter process is applied with galaxy repetitions, which are required to conserve galaxy weights -- without galaxy repetitions (no repeats), shown in grey, the distributions are discrepant. Adding a jitter to the positions of galaxies is able to mask the projected effect of fibre collisions on the distribution of galaxies.}
\end{figure*}

\subsubsection{Masking small scale survey systematic effects}

All galaxy surveys will suffer from small scale systematic effects; for photometric surveys these are typically introduced due to the blending of galaxies in crowded fields, while for spectroscopic surveys fibre collisions limit the number of spectra that can be obtained for galaxies with close neighbours on the sky. In either case, we can no longer trust the distribution of galaxies on scales below these angular scales. This is slightly different to the difference in the LF and ALF distributions discussed previously, as in this scenario the systematic occurs on the projected distribution of galaxies on the sky. This means galaxies that are close by on the sky but actually quite distant can still suffer from these systematic effects.

To understand the consequences of systematic effects on our measurements we take galaxies from the Millennium XXL octant described in section~\ref{sec_galcat} and iteratively remove galaxies with an angular separation of $< 30\,{\rm arcsec}$. For each galaxy with more than one close pair (i.e. a galaxy with angular separation $< 30\,{\rm arcsec}$) we first match galaxies with their close pairs and randomly select a galaxy to keep. To ensure the missing galaxies are accounted, we add the weights of the removed galaxies to the weight of galaxy being kept. Each galaxy now has either a weight of zero (if it has been removed), a weight of one (if it has no close pairs or no pairs were lost), or an integer weight greater than one (to account for the lost pairs). This gives us a catalogue of galaxies with an approximate fibre collision-like systematic, which we will refer to as the no close pair (No CP) catalogue.

We measure the 2PCF and MST on the Millennium XXL octant with all galaxies and with No CP shown in dark blue in Fig.~\ref{fig_jitter_nocp}. For the No CP catalogues, galaxies with integer weights greater than one are repeated by the weighted number in the catalogue. This ensures they are correctly weighted in the 2PCF, however for the MST, this step has practically no effect on the results, since this introduces edges of length zero in the MST, which has no physical meaning and is no different to the results we would obtain if galaxies were not replicated. Nevertheless we can see that removing the close pairs in the 2PCF reduces the amplitude of small scale clustering, although this remains consistent with a $\chi^{2}_{r}=0.76$. For the MST, the distribution of $l$ and $b$ are significantly different showing $\chi^{2}_{r}=88.2$ for $l$ and $\chi^{2}_{r}=17.8$ for $b$. These distributions are significantly different and highlights the effect that fibre collisions can have on the MST and other hard-wired algorithms. 

We now remeasure the 2PCF and MST with a jitter, defined with a jitter dispersion $\sigma_{\rm J}=3\,h^{-1}{\rm Mpc}$, shown in light blue in Fig.~\ref{fig_jitter_nocp}. This removes any discrepancy in the MST statistics between the original and No CP catalogues, most significantly reducing $\chi^{2}_{r}=88.2$ for $l$ to $\chi^{2}_{r}=0.41$. The results show that adding a jitter to the position of points is able to remove the systematic effects caused by fibre collisions. Crucially, it removes a problem caused by galaxies with weights $>1$. In the MST this inevitably leads to edges equal to zero. With a jitter, this problem is removed entirely, since the repetition of galaxies with a weight $>1$ will never lie exactly on top of each other. This ensures the missing galaxies are correctly accounted for in the MST and indeed any hard-wired statistic. 

\begin{figure*}
    \centering
    \includegraphics[width=\textwidth]{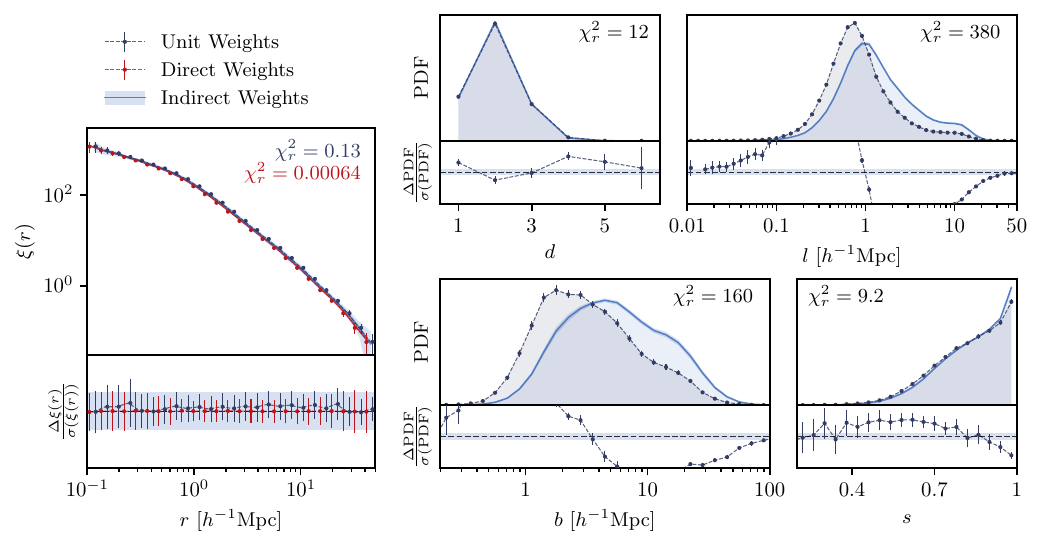}
    \caption{\label{fig_invdens}We measure the 2PCF (left) and MST statistics for the Millennium XXL galaxy catalogue with (light blue) and without (dark blue) inverse density weights, imposed via Monte Carlo subsampling. The Monte Carlo procedure imposes inverse density weights on the galaxy distribution at the catalogue level. The inverse density weights homogenise the distribution, and removes the redshift selection function systematic from the measurements. The 2PCF shows the Monte Carlo procedure is able to impose the correct weights, the 2PCF with and without weights are completely consistent with a $\chi^{2}_{r}=0.13$. Note, for consistency the randoms are assigned inverse density weights directly in the Monte Carlo 2PCF measurement. On the otherhand, the MST distributions of $d$, $l$, $b$ and $s$ show strong deviations with and without the Monte Carlo procedure -- up to $\chi^{2}_{r}=380$ for $l$, demonstrating how different the MST can be if galaxies are used directly with the MST without accounting for their weights and the survey's redshift selection function.}
\end{figure*}

In practice the jitter dispersion scale will need to be calibrated to be large enough to ensure sensitivity to small scale systematic effects like fibre collisions are removed. In this current setup the No CP catalogue suffers from percolation effects since close pairs percolate due to the algorithm used to construct them, this is one way the approximation departs strongly from the real effect of fibre collisions. In practice we would estimate the jitter dispersion scale by comparing mock datasets with and without fibre collisions as a function of $\sigma_{\rm J}$ and finding the smallest value that ensures consistency.

\subsection{Monte Carlo subsampling performance}

We test the performance of the Monte Carlo subsampling scheme, discussed in section~\ref{sec_monte_carlo_prob}, to assign galaxy weights and mitigate survey selection functions at the catalogue level. This is used to determine the weighted hard-wired statistics without needing to understand how to pass weighted points directly to the hard-wired algorithm itself.

\subsubsection{Inverse redshift selection weights}

Galaxy surveys observe galaxies inhomogeneously and anisotropically, effects induced by the galaxy survey's footprint and redshift selection function. The redshift selection function forces inhomogeneities in the dataset which needs to be accounted for in any cosmological measurement, as we do not want to confuse properties of the survey with cosmology. For the 2PCF and indeed most $N$-point correlation functions this is readily handled by including a catalogue of randoms which captures the survey's footprint and selection function. This aptly removes any sensitivity in the 2PCF to the survey properties. This is not the case for the MST and any hard-wired algorithms, meaning results will be the product of cosmology and the survey's footprint and selection functions.

To address this we assign galaxies additional weights that re-weight galaxies based on the inverse of the redshift density
\begin{equation}
    w_{\rm g}=\frac{\rho_{\rm target}}{\rho(z_{\rm g})},
\end{equation}
where $\rho(z_{\rm g})$ is the redshift selection function density for a galaxy at redshift $z_{\rm g}$ and $\rho_{\rm target}$ the target density for subsampling galaxies (set to $\rho_{\rm target}=0.001$ in this analysis).

We will first concentrate on the 2PCF and the performance of the weight assignment via Monte Carlo subsampling. The 2PCF is calculated with a random catalogue with 10 times the density of the galaxy catalogue. The 2PCF is first computed with unit weights, i.e. treating each point equally, indicated in Fig.~\ref{fig_invdens} in dark blue. We then compute the 2PCF with inverse density weights applied directly in red and indirectly, using the Monte Carlo subsampling procedure computed over a 100 iterations, in light blue. Note, for the randoms we directly assign them inverse density weights in the 2PCF to account for the imposed homogeneity of the distribution being sampled via Monte Carlo subsampling. The 2PCF in Fig.~\ref{fig_invdens} shows: (1) weighting galaxies and randoms by the inverse of the density (where the density here refers to density caused by selection functions and varying completeness) has no effect on the 2PCF, since randoms are designed to remove any systematic effects from the survey's footprint and selection function;
and (2) the Monte Carlo subsampling procedure is able to impose weights on the galaxy distribution at the catalogue level and retrieve the directly weighted 2PCF with a $\chi^{2}_{r}=0.00064$.

\begin{figure}
    \centering
    \includegraphics[width=\columnwidth]{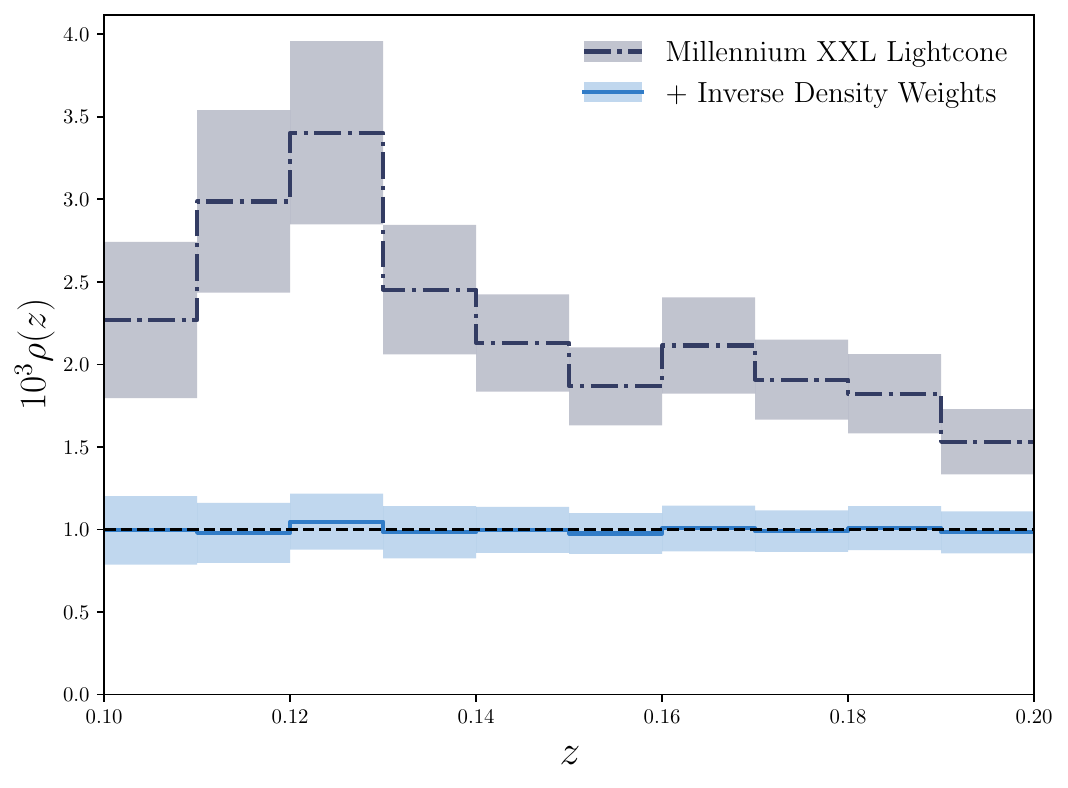}
    \caption{\label{fig_nz}The redshift selection function density $\rho(z)$ for the Millennium XXL galaxy lightcone with (in dark blue) and without inverse density weights (in light blue). The dashed line shows the target density $\rho_{\mathrm{target}}=0.001$. Inverse density weights produce galaxy sub-samples with the intended homogeneous redshift distribution.}
\end{figure}

\begin{figure*}
    \centering
    \includegraphics[width=\textwidth]{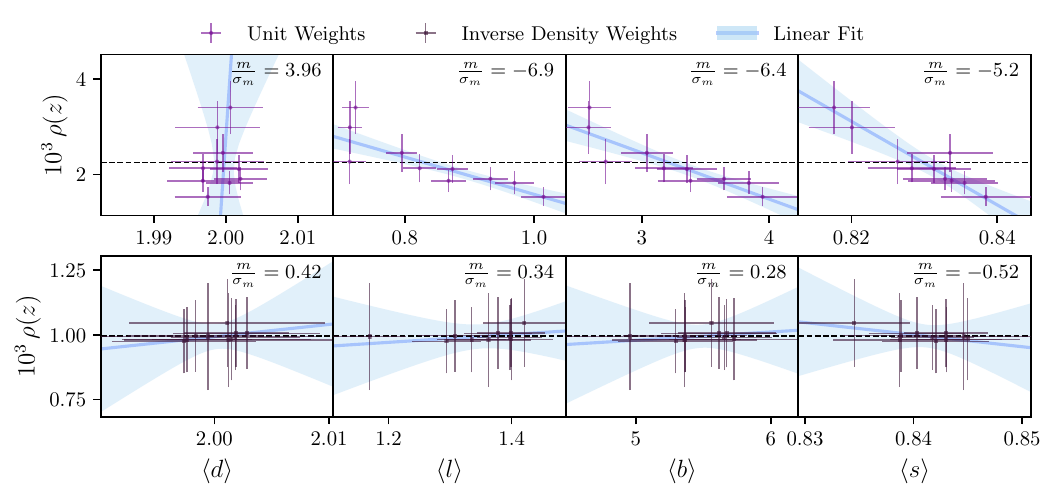}
    \caption{\label{fig_mst_vs_nz}The mean of the MST statistics is shown as a function of the redshift selection function $\rho(z)$ for the Millennium XXL galaxy catalogue. Using unit weights (top panel) the MST shows strong correlations with $\rho(z)$. The significance of the gradient $m$ of the linear function fitted to the data is indicated on the top right of the subpanels. For the MST without weights the significance of the gradient is $|m/\sigma_{m}|\gtrsim 4$, meaning there are strong correlations. In the bottom panel we compare the MST statistics after applying inverse density weights, via Monte Carlo subsampling. Here we show that the MST statistics have no correlation with the selection function, with the significance of the gradient measured to be $|m/\sigma_{m}|\lesssim 0.5$.}
\end{figure*}

The MST distributions with (light blue) and without (dark blue) the inverse density weighting scheme is shown in Fig.~\ref{fig_invdens} on the right. The figure demonstrates how biased the MST can be if the survey's selection function is not accounted for in the measurements. By imposing the inverse density weights we homogenise the distribution at the catalogue level before making the MST measurement (see Fig.~\ref{fig_nz}). This ensures the statistic is no longer dependent on the survey's redshift selection function, which is demonstrated in Fig.~\ref{fig_mst_vs_nz}.  We plot the mean of the MST statistics as a function of the redshift selection function in ten redshift slices between $0.1 \leq z \leq 0.2$ and plot this against the Millennium XXL density as a function of redshift, $\rho(z)$. The top panels shows the relation with unit weights, while the bottom panels show the relation when using Monte Carlo subsampling to apply inverse density weights. We fit a linear relation to the points and show that without weighting the MST the statistics is strongly correlated to the redshift density of the distribution, this is not surprising since we expect the MST to be extremely sensitive to density. Using the inverse density weights, this sensitivity is completely removed and the MST statistics are no longer correlated with the redshift selection function.

\begin{figure}
    \centering
    \includegraphics[width=0.925\columnwidth]{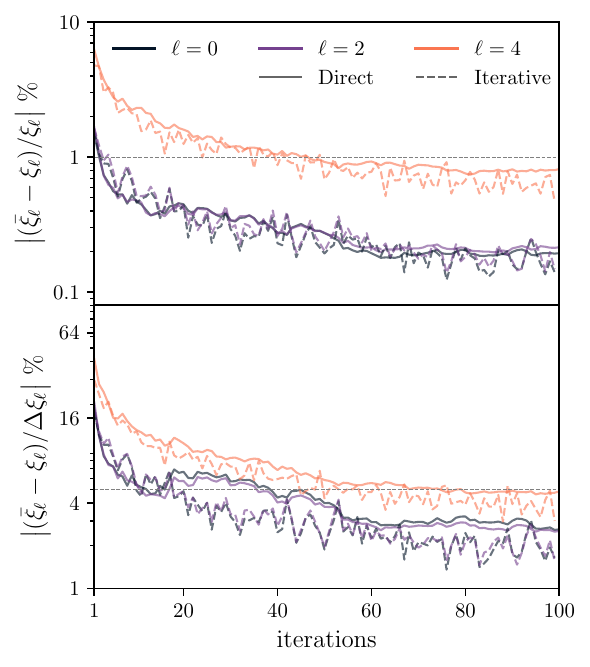}
    \caption{\label{fig_error_xi}The convergence of the Monte Carlo subsampling is shown for the 2PCF multipoles (the monopole $\ell=0$, the quadrupole $\ell=2$ and hexadecapole $\ell=4$). The mean fractional difference (top panel) and mean significant difference (bottom panel) are shown as a function of Monte Carlo iterations. In solid lines we show the direct method where we use the true weighted 2PCF multipoles to compute the convergence and in dashed lines we use the iterative estimator. The iterative method is shown to be completely consistent with the direct method.}
\end{figure}

\subsubsection{Convergence and signal-to-noise}

The Monte Carlo subsampling procedure allows us to place galaxy weights, for mitigating survey systematic effects, at the catalogue level. The iterative procedure allows us to indirectly apply the weights to the statistics of interest. In Sec.~\ref{sec_convergence_criteria} we introduce several convergence criteria that can be used to determine the number of Monte Carlo iterations $n$ required to make the measurement to the desired accuracy.

The mean fractional difference $\epsilon_{f}$ gives us a measure of the error of the Monte Carlo statistics in relation to the true weighted statistics. In most cases we will not have access to $\epsilon_{f}$, but instead only have access to $\epsilon_{I}$ which is computed from the sequential iterations of the statistic. This is extremely useful in cases where errors are difficult to obtain and the convergence criteria can be achieved by requiring the measurement be made to within $1\%$ error. In the cases where we do have access to the error in the measurement we can use the mean significant difference $\varepsilon_{f}$.

We test the accuracy of the iterative estimators ($\epsilon_{I}$ and $\varepsilon_{I}$) to compute $\epsilon_{f}$ and $\varepsilon_{f}$ respectively, using the 2PCF where galaxy weights can be assigned directly. In Fig.~\ref{fig_error_xi} we compare the convergence estimators for the 2PCF multipoles ($\ell=0$ the monopole, $\ell=2$ the quadrupole and $\ell=4$ the hexadecapole). The figure shows the direct estimator, using the true weighted multipoles, in relation to the iterative estimators, using the Monte Carlo procedure. The iterative estimator follows the distribution for the direct estimator of the mean fractional difference (top panel) and the mean significant difference (bottom panel). This means we can rely on the iterative estimator to determine the necessary iterations required for estimating the weighted MST and other hard-wired statistics when we use Monte Carlo subsampling.

\begin{figure}
    \centering
    \includegraphics[width=0.925\columnwidth]{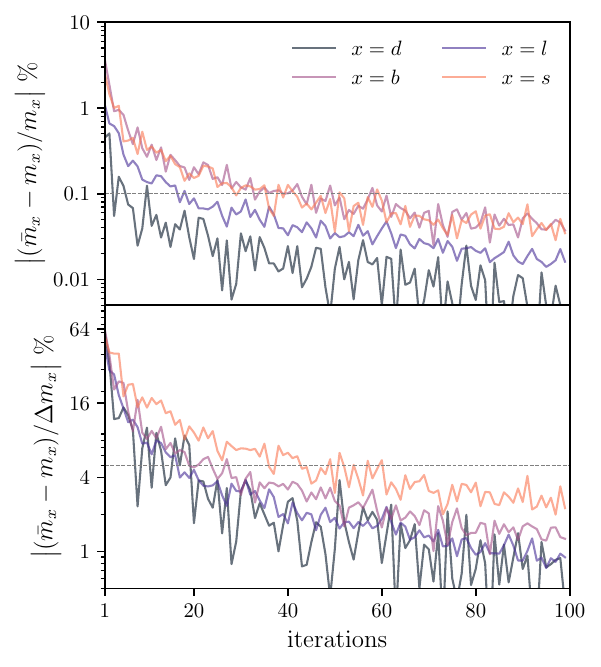}
    \caption{\label{fig_error_mst}The convergence of the Monte Carlo subsampling is shown for the MST distributions using the iterative estimator. In the top panel we show the mean fractional error and in the bottom panel we show the mean significant error as a function of iterations.}
\end{figure}

In Fig.~\ref{fig_error_mst} we compute the mean fractional difference and mean significant difference for the MST distributions as a function of Monte Carlo iterations. These are computed using the iterative estimator since weights cannot be applied directly to the MST. Here we illustrate how the MST statistics converge, showing that in general the degree is the quickest to converge followed by the edge length, branch length and then branch shape. Some aspect of this is dependent on the bin sizes used for the MST distribution measurements. In the case where iterations are prohibitively expensive, the user could increase the bin sizes to reduce the iterations required for convergence. The values of $\epsilon_{f}$ and $\varepsilon_{f}$ allow us to know how accurate our estimates of the weighted MST are, for example we know from $\epsilon_{f}$ (top panel) that we can achieve a $0.1\%$ measurement of the MST statistics after $\gtrsim 50$ iterations and an error of $5\%$ as a fraction of the standard deviation.

\begin{figure}
    \centering
    \includegraphics[width=0.95\columnwidth]{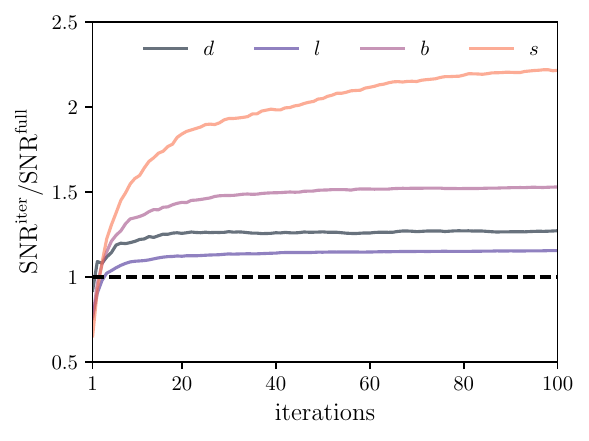}
    \caption{\label{fig_SNR}The signal-to-noise ratio (SNR) of the MST distribution measured from the iterative estimator ($\rm SNR^{iter}$) with inverse density weights is compared to the SNR of the MST performed on all galaxies ($\rm SNR^{full}$). The SNR with the iterative estimator is shown to be larger for all MST distributions after a few iterations. This is because the inverse density weights allow us to marginalise over the variation in the redshift selection function, making the measurement a more well defined measure of the MST at a specific galaxy density.}
\end{figure}

We measure the mean SNR ratio of the MST
\begin{equation}
    {\rm SNR} = \frac{1}{n_{\rm d}}\sum_{i=1}^{n_{\rm d}}\frac{m_{x}}{\Delta m_{x}},
\end{equation}
where $m$ is the MST measurement/data vector, $\Delta m$ the associated standard deviation, and $x$ is used to denote the MST statistic $d$, $l$, $b$ or $s$. The ${\rm SNR}$ is measured for the MST distributions measured on all the galaxies (${\rm SNR}^{\rm full}$) and then the iterative MST distributions measured with the Monte Carlo procedure (${\rm SNR}^{\rm iter}$). In Fig.~\ref{fig_SNR} we compare the SNR of the full MST in relation to the iterative estimator using inverse density weights. The plot shows that in all cases the SNR ratio of the measurement is improved by marginalising over the redshift selection function with inverse density weights. A concern in introducing inverse density weights and homogenising the distribution before applying the MST, is that this process loses information and reduces the information content of the data. This concern only turns out to be true for a few iterations, after more than a few iterations the SNR shows the opposite effect. Marginalising over the redshift selection function we improve the SNR and are therefore extracting more information from the data. The improved SNR arises from the fact that our measurement is more well defined, it is a measure of cosmology with the effect of the survey's selection function limited, while prior to this the MST measurement was a mixture of both cosmology and the survey's selection function, diluting the information of interest to study cosmology.

\begin{figure*}
    \centering
    \includegraphics[width=\textwidth]{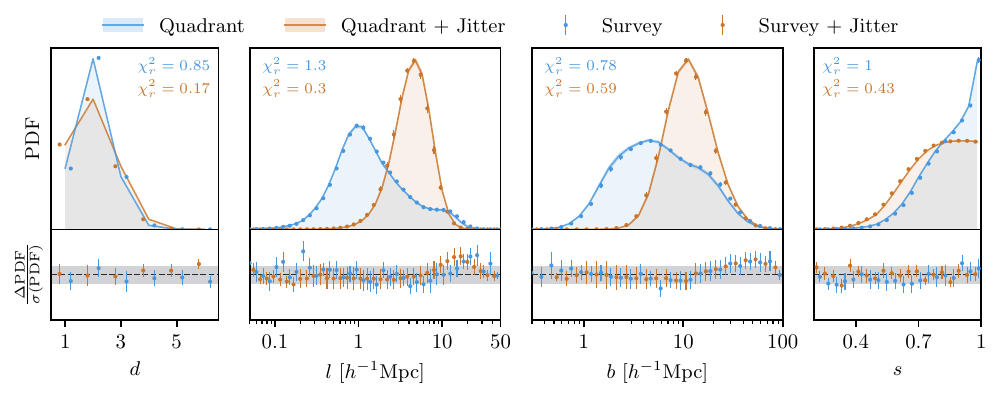}
    \caption{\label{fig_mst_boundary}The octant to survey bias is illustrated for the MST distributions with and without a jitter. This shows the bias caused by placing galaxies into a realistic survey footprint -- using the BOSS LOWZ North footprint. For the degree the bias is completely negligible, while for the edge length and branch length we see a raised bump at higher values of $l$ and $b$, most prominent for the measurements made with a jitter. For $s$ the significance of the bias increases with a jitter but has no characteristic shape, suggesting the measurement may be too noisy to distinguish this clearly.}
\end{figure*}

\subsubsection{Survey footprint bias}

The footprint for a galaxy survey is dictated by the location of the observations (if on the ground), galactic and zodiacal foregrounds, bright stars and observing conditions. This will create a footprint which is unique to the survey, non-trivial and sometimes difficult to replicate. For statistics that require $N$-body simulations, we need to decide whether we want to emulate the MST or hard-wired algorithm with the survey footprint already superimposed on the distribution of galaxies or to emulate it in a topology more readily accessible for simulations, such as a cubic box or for a lightcone on a octant on the sky. The benefit of the latter, is the measurement is independent of the survey and can be ported to multiple different surveys. In all cases to use the emulated statistic and perform inference with galaxy survey measurement requires learning the footprint bias -- i.e. the bias that the footprint imposes on the measured statistic. We will assume the bias is small and can be learnt from mocks produced at a fiducial cosmology imposing the survey's footprint -- in practice this may not be true, in which case it may be beneficial to emulate the statistics with the survey footprint already imposed on the simulation data.

In Fig.~\ref{fig_mst_boundary} we make measurements of the MST on the Millennium XXL galaxy catalogue in a octant on the sky and with the BOSS LOWZ North survey imposed (simply referred to as the `survey'). The statistics are measured with inverse density weights to measure the MST with a galaxy density of $0.001\, h^{3}\,{\rm Mpc}^{-3}$ and shown with and without a jitter with dispersion $\sigma_{\rm J}=5\,h^{-1}{\rm Mpc}$. The measurements show that for the MST statistics, the footprint bias from a octant to the LOWZ North survey mask is small, and is smaller after applying a jitter. Since the footprint bias is small, the functional form of this bias can be learned by applying the MST to a galaxy simulation in a cubic box or octant used for the rest of the simulations and then applying it to the same simulation with the survey footprint imposed. This bias can then be subtracted from the observational measurements or added to the theoretical predictions from simulations prior to parameter inference.

\begin{figure*}
    \centering
    \includegraphics[width=\textwidth]{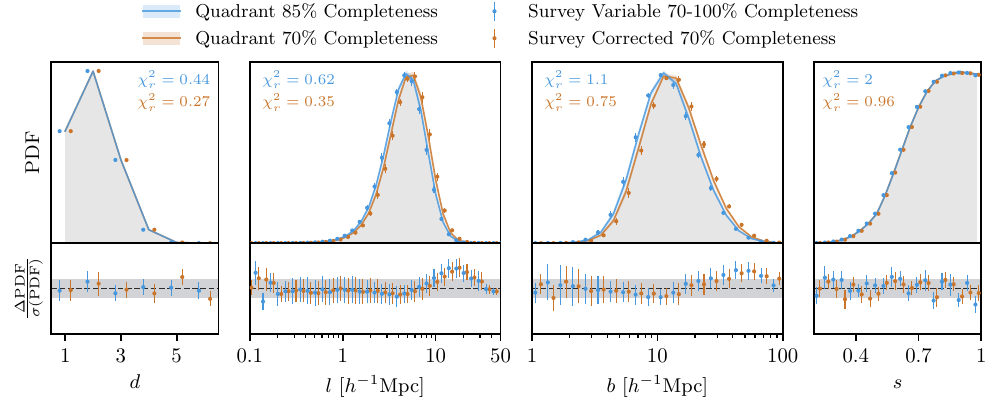}
    \caption{\label{fig_mst_completeness}The MST distributions (left to right: degree $d$, edge length $l$, branch length $b$, and branch shape $s$) are shown for the Millennium XXL galaxy catalogue placed in a octant and analyzed with varying completeness levels (ranging from 70\% to 100\%) within the LOWZ survey footprint. In blue, we compare the octant with 85\% completeness and variable survey completeness, while in orange, we compare the octant with the survey corrected to 70\% completeness. The figure demonstrates that using variable completeness provides a reasonable approximation of the MST measurements, but correcting for completeness significantly reduces discrepancies between the two measurements. The most prominent remaining discrepancies are attributed to biases from the survey footprint.}
\end{figure*}

\begin{figure}
    \centering
    \includegraphics[width=\columnwidth]{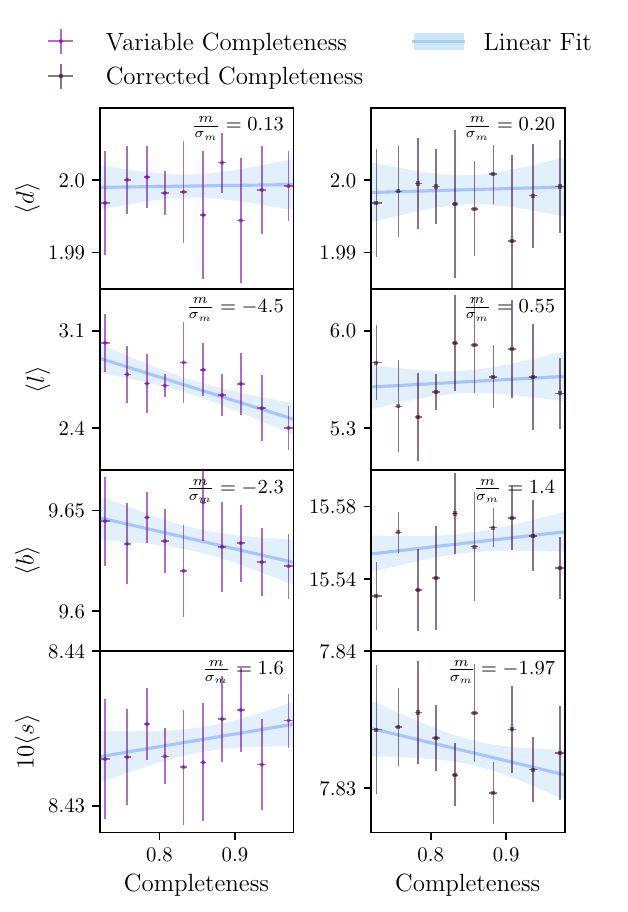}
    \caption{\label{fig_compvsmst}The mean of the MST distributions is plotted as a function of completeness in the survey. On the left we plot the relation with variable completeness and on the right we plot the relation after correcting for completeness. A linear fit is constructed from the measurements, shown in blue with $2\sigma$ confidence envelopes. With variable completeness we see a clear anticorrelation between $l$ and $b$ with the significance of the gradients $|m/\sigma_{m}| > 2$. After correcting for the completeness, all the linear fits are consistent with no correlation, with $|m/\sigma_{m}| < 2$ for all statistics. The remaining discrepancies are likely caused by variations in the survey footprint caused by segmenting the sky according to completeness.}
\end{figure}

\subsubsection{Inverse angular selection weights}

Galaxy surveys will generally aim to observe the sky up to a certain depth across the survey's footprint. However, seeing conditions, instrumentation and foregrounds can limit the degree to which this can be achieved. This will create variations in the number of galaxies observed across the sky, which is often characterised by a measure of completeness. If unaccounted, the variation in galaxies observed across the sky will be a systematic that can alter our results and could lead to biased or incorrect parameter inference. It is therefore important that this effect is carefully included and mitigated.

To test the effect of completeness, in Fig.~\ref{fig_mst_completeness}, we subsample galaxies from the LOWZ North footprint with a completeness fraction that varies linearly in RA, from 1 to 0.7. This is quite an extreme set up designed to illustrate how such a systematic could bias measurements, but not one we expect to see in real data. We measure the MST distribution for the survey with varying completeness and compare it to a measurements made on the octant with a completeness fraction of 0.85 (shown in blue). The measurements are made with inverse density weights used for the radial selection function and a jitter with dispersion $\sigma_{\rm J}=5\,h^{-1}{\rm Mpc}$. When we apply the MST to the galaxies with varying completeness we see a more significant offset in relation to a octant with an effective completeness of $0.85$ (i.e. the mean of the variable completeness of the survey). We then apply inverse completeness weights, making full use of Eq.~\ref{eq_probgal}, setting $C_{\rm target}=0.7$. This homogenises the distribution of galaxies across the sky, prior to the computation of the MST. This means the completeness systematic does not enter the MST statistics. For comparison we subsample the octant with a completeness fraction of 0.7 (shown in orange). The relations show that correcting for discrepancies in completeness reduces the $\chi^{2}_{r}$. The main discrepancies that still remain are characteristic of the footprint bias shown in Fig.~\ref{fig_mst_boundary}, which have not been removed in this analysis.

In Fig.~\ref{fig_compvsmst} we measure the mean of the MST statistics as a function of completeness. Here we use the average completeness in jackknife tiles on the simulated survey. We find that the mean of the distributions for $l$ and $b$ show a clear relation with respect to completeness, where the significance of the linear gradient is $|m/\sigma_{m}|> 2$. After applying the completeness correction, the linear relation with completeness is greatly reduced, with $|m/\sigma_{m}|<2$.

\begin{figure*}
    \centering
    \includegraphics[width=\textwidth]{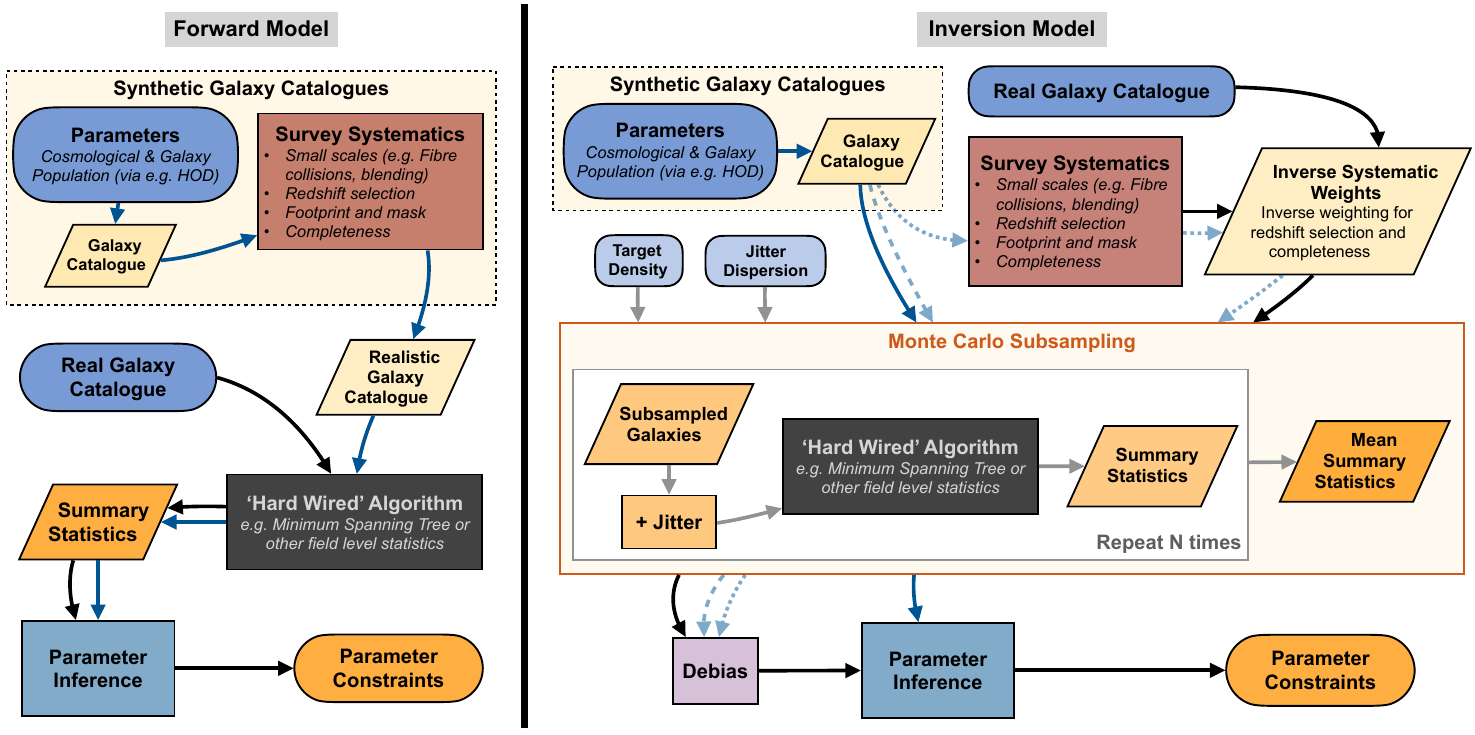}
    \caption{\label{fig_schematic}A schematic diagram of a traditional forward model parameter inference pipeline for the MST or hard-wired algorithm (on the left) is compared to the inversion model approach being proposed in this paper (on the right). Where the two methods strongly diverge is in the generation of synthetic data. In the forward model approach, survey systematic effects are imprinted onto the galaxy catalogues and fed into the resulting data vectors, while in the inversion model approach Monte Carlo subsampling and a jitter are used to remove survey systematic effects from the measurements on real data and synthetic data are used to make predictions at a specific target density. The resulting data vectors for the inversion model are purely a function of cosmology and galaxy population physics. In the inversion model we illustrate with blue transparent arrows an extra validation step where a single synthetic galaxy catalogues is injected with survey systematic effects to validate the inversion methodology and compute any biases that may remain.}
\end{figure*}

\section{Discussion}
\label{sec_discussion}

The MST and other hard-wired field level statistics offer an alternative probe of large scale structure, capturing higher-order statistical information present in the field and cosmic web which are not generally attainable from measuring the 2PCF. However, we are yet to capitalise on the additional information in these statistics because making these measurements precisely on real data is fraught with observational and theoretical systematic effects and uncertainties.

The conventional approach for dealing with hard-wired algorithms such as the MST is to make forward modelled predictions from $N$-body simulations. The simulations would vary both cosmological parameters and galaxy population parameters (such as HODs) and forward model survey systematic effects. This requires synthetic galaxy catalogues with:
\begin{enumerate}
\item Galaxy weights -- used to forward model the galaxy weight bias caused by ignoring or, effectively, using unit weight (i.e.~setting all galaxy weights to one) in the MST or hard-wired algorithm.
\item Small scale systematic effects -- such as fibre collisions and systematic effects from crowding need to be injected since the MST and some hard-wired algorithms may not be able to impose small scale cuts.
\item Redshift selection function -- that may be challenging to impose if the target selection is complex.
\item Footprint and angular selection functions -- that may not be trivial if they are dictated by observational systematic effects.
\item Galaxy bias evolution -- galaxies observed at higher redshift will often be more luminous and more massive, this means our tracer for large scale structure changes as a function of redshift. The evolution of galaxy bias may need to be forward modelled.
\end{enumerate}
While it is common to reproduce all of these systematic effects in mocks for computing covariances \citep[see][]{Kitaura2016, Ereza2024}, reproducing these on synthetic data for parameter inference is much more challenging due to the complexity of the task and the number of simulations utilised. Although all these requirements may be satisfied, we cannot be certain we can trust the small scale physics from $N$-body simulations. Similarly, even in cases where these are produced from hydrodynamic simulations, we cannot be certain that the galaxy formation and evolution prescriptions are completely accurate. These issues mean our predictions from the MST and hard-wired algorithms may be fatally flawed. Furthermore, forward modelling will need to be carried out independently for every galaxy survey, due to the specific nature of survey systematic effects. This means the analysis can only be compared to other surveys at the parameter level, since measurements made are a measure of cosmology, galaxy population and systematic effects -- the latter of which will be unique to the survey and will make comparisons between surveys practically impossible. This will become important if the constraints on cosmological parameters challenges the standard model, requiring deeper investigation and interpretation.

The methods in this paper try to resolve these issues by showing how to incorporate galaxy weights, mitigate small scale uncertainties and galaxy survey systematic effects to produce measurements from surveys that are solely dependent on cosmology and the galaxy tracers being used. This simplifies the requirements from $N$-body simulations which will no longer be required to reproduce small scale systematic effects like fibre collisions, and galaxy survey redshift and angular selection functions. This means emulators of the MST statistics or hard-wired algorithm can be built more generally, substantially reducing the computational requirement for parameter inference. Furthermore, it will allow the measurements to be more readily interpreted, since the measurement is no longer survey dependent due to survey specific selection functions and systematic effects. The only survey specific issue that will remain is the footprint bias, which is the bias in making a measurement in the survey's footprint in comparison to making it in an $N$-body box or lightcone octant (or in some cases across the full-sky). The footprint bias can be computed by superimposing synthetic mock galaxy data into the survey's footprint and can then be removed from the survey measurement or injected into the predictions depending on the size and nature of the bias.

To implement this method we rely on two concepts, jittering and Monte Carlo subsampling. A jitter is the process of adding noise to the positions of galaxies in data, in our analysis the noise added follows a Gaussian distribution with a user specified dispersion scale $\sigma_{\rm J}$. Jittering allows us to masks small scale uncertainties from theoretical uncertainties in simulations and small scale systematic effects from observations, such as fibre collisions. The jittering process is the point-process equivalent to smoothing a field. The method is shown to be able to remove small scale theoretical differences, as well as fibre collision-like systematic effects in a mock galaxy catalogue. Monte Carlo subsampling is the process by which we can apply weights to galaxy catalogues indirectly when the algorithm or statistics being measure, like the MST, are unable to include weighted points. Monte Carlo subsampling treats galaxy weights as probabilities, drawing galaxies based on their probabilities and performing the MST or hard-wired algorithm on the subsampled galaxies iteratively. The mean of the statistics computed from the subsampled galaxies is taken to be the weighted statistics. We demonstrate the subsampling procedure applies weights correctly by applying the technique to the 2PCF where weights can be applied directly. We also show that without this indirect application of weights, the measurements of the MST are extremely biased with a reduced $\chi^{2} \approx 400$ in some cases.

We extend the Monte Carlo subsampling to remove survey dependent selection functions, such as the redshift selection function and completeness. To achieve this, we adjust the weights of galaxies by the inverse of the redshift selection function and completeness. This allows us to draw subsampled galaxies uniformly across the survey at some specified target density. This method removes any dependence between the survey's selection functions (redshift and completeness) and the MST or hard-wired statistics measured.

This approach to measuring the MST, and indeed any hard-wired algorithm, has a number of advantages over the traditional forward model approach. 
\begin{itemize}
    \item Jittering and Monte Carlo subsampling occur at the catalogue level, meaning they can be relied upon for any statistics, allowing for galaxy weights to be applied correctly and survey systematic effects to be largely removed.
    \item The method targets a measurement at a specified target galaxy density. This is a better defined measurement then simply making the measurement on all galaxies in the survey and leads to measurements with higher SNR ratios.
    \item Measurements are not survey specific and can readily be compared to at the data vector level rather than purely at the parameter level. This will greatly improve interpretability and reliability, especially in cases where the inference from the MST or hard-wired algorithm challenges the standard model.
    \item It removes the heavy cost of forward modelling and enables emulators to be built for a more general use rather than being custom built for a single galaxy survey.
\end{itemize}

In this analysis we have limited our analysis to galaxies with stellar masses greater than $10^{13}\,h^{-1} M_{\odot}$. In practice how galaxies are limited in a cosmological analysis could play a significant role in the complexity of the galaxy population modelling. A volume limited sample, i.e. a complete sample of galaxies above a certain luminosity or mass, will make modelling significantly easier. This removes the need to model Malmquist bias and incomplete galaxy populations at the low mass end.

In Fig.~\ref{fig_schematic} we compare the schematic approach of a traditional forward modelled parameter inference pipeline in comparison to the inversion model being proposing in this paper. The key difference in the approaches is that the inversion model removes the impact of survey systematic effects on the MST or hard-wired algorithm on real data, while the forward model approach adds survey systematic effects to synthetic data.

We outline below the steps required to make measurements of the MST or hard-wired statistic using the inversion model and consequently how to infer parameter constraints from simulations.
\begin{enumerate}
    \item Measurements from galaxy survey:
    \begin{enumerate}
        \item Define the properties of the galaxy population to compute the MST or hard-wired statistic. This could mean creating a volume limited sample of galaxies in mass or luminosity. Note, the more complex the properties the more complex the galaxy population modelled will need to be.
        \item Measure the redshift selection function on the galaxy population.
        \item Assign galaxies inverse density weights to remove the redshift selection function and variations in completeness across the sky.
        \item Using mocks with and without small scale systematic effects (like fibre collisions) determine the jitter dispersion  $\sigma_{\rm J}$ required to mask small scale systematic effects. This is determined by finding the smallest value of $\sigma_{\rm J}$ which produces consistent measurements with and without small scale systematic effects.
        \item Determine the number of Monte Carlo subsampling iterations required for the data vector to converge.
        \item Test measurements are not correlated with systematic effects, to test systematic corrections have been correctly applied.
    \end{enumerate}
    \item Measurements from $N$-body simulations:
    \begin{enumerate}
        \item Model galaxy population and perform similar cuts to real data.
        \item Assign galaxies subsampling weights to make measurements at the correct effective galaxy density.
        \item Determine the number of Monte Carlo subsampling iterations required for the data vector to converge.
        \item At a fiducial cosmology and galaxy population model, compute the footprint bias of the survey by making the measurement in the original simulation box or lightcone octant and then imposing the survey’s footprint on the sample.
    \end{enumerate}
    \item Validation:
    \begin{enumerate}
        \item Using a single synthetic catalogue, test that survey systematic effects are correctly removed through the inversion pipeline.
        \item Test predictions from N-body simulations with the same cosmological parameters and galaxy population modelling match measurements from galaxy survey mocks.
        \item Test the parameter inference pipeline is unbiased by constraining cosmological parameters and galaxy population models on mock observations.
    \end{enumerate}
    \item Apply parameter inference pipeline to real data.
\end{enumerate}

It is important to note, in all cases we assume the $N$-body measurements are made in redshift space, either using a lightcone or by adding redshift space distortions along a chosen line-of-sight axis inside an $N$-body periodic box. However, for the MST the measurements are made in comoving distance, where redshifts are converted into comoving distances assuming a fiducial cosmology. For simulated lightcones a like-for-like analysis can be performed because this conversion can be carried out consistently however for periodic boxes this conversion becomes problematic. Since stretching the box to carry out the conversion to the fiducial comoving distance will alter the footprint bias. For this reason it may be simpler to project the periodic box across a region on the sky, so that the positions can be converted into redshift and reprojected into the fiducial comoving distance. To ensure the footprint bias is the same we would limit the analysis to the redshift ranges of the survey.

\section{Summary}
\label{sec_summary}

In this paper, we outline how to make precision measurements of the MST from galaxy surveys. The technique relies on two methods: jittering and Monte Carlo subsampling. 

Jittering is a point-process smoothing technique, where the positions of points are given a random `jitter' or noise, that follows a Gaussian with a standard deviation given by the jitter dispersion $\sigma_{\rm J}$. In Fig.~\ref{fig_jitter_LF} we show how jittering can remove the small scale differences of two random walk distributions -- illustrating that jittering can mask theoretical small scale uncertainties. In Fig.~\ref{fig_jitter_nocp} we show how jittering removes the effects of fibre collisions -- illustrating that jittering can mask small scale systematic effects from galaxy surveys.

Monte carlo subsampling is a technique for indirectly incorporating galaxy weights to the MST. The MST is a hard-wired algorithm, meaning its internals and outputs cannot be altered or post-processed in a consistent way. This presents a unique challenge for galaxy surveys, as there are no methods for directly including galaxy weights to the MST. In Monte Carlo subsampling, galaxy weights are treated as probabilities for sampling. The MST is then performed on subsampled realisations of galaxies. Taking the mean of the MST distributions over many realisations gives us the MST of the weighted galaxies. In Fig.~\ref{fig_invdens} we show this technique can reproduce the weighted 2PCF and show that failing to include galaxy weights can deeply bias the MST. Furthermore, we show how to alter the weights of galaxies to correct for variations in the survey's redshift selection function and completeness across the sky. The fundamental change in this approach is that we target a measurement of the MST for galaxies at a specified target density. This ensures the MST has no correlations with the redshift selection function (see Fig.~\ref{fig_mst_vs_nz}) and completeness (see Fig.~\ref{fig_compvsmst}). In Sec.~\ref{sec_convergence_criteria} we derive measurements of convergence which can be used to determine the number of iterations required for Monte Carlo subsampling. We illustrate the convergence as a function of iterations for 2PCF using a direct and indirect approximation for the convergence in Fig.~\ref{fig_error_xi}, and for the MST using the approximation in Fig.~\ref{fig_error_mst}. In Fig.~\ref{fig_SNR} we show that Monte Carlo subsampling improves the SNR of the MST measurements.

In Fig.~\ref{fig_mst_boundary} we illustrate how the technique allows us to compare predictions from measurements made with quite different footprints on the sky. The only thing that needs to be corrected is the survey's footprint bias, which is the bias in making measurements in the survey's footprint in comparison to making it in a simulation (periodic box or lightcone octant). This is a bias that can be learnt and corrected.

Lastly, in Fig.~\ref{fig_schematic} we illustrate how a parameter inference pipeline will need to be altered to include this inversion method and in section~\ref{sec_discussion} we outline the steps that need to be taken to make a measurement and infer cosmological parameter from the MST. The technique introduced in this paper resolves a number of challenges faced by many hard-wired field level statistics, not only the MST, and because the techniques operate at the catalogue level, they can be generally used for any statistics. By mitigating survey systematic effects we can make measurements that can now be readily compared between surveys and can rely on emulators and simulations that are not custom built for a single survey. This will improve interpretability and will make it easier for us to reap the rewards and sensitivities of field level statistics from the next generation of galaxy surveys, such as DESI, \emph{Euclid}, LSST and WST.

\section*{Acknowledgements}

We thank Benjamin Joachimi, Nicolas Tessore, Tessa Baker, Shaun Cole and Grazianno Rossi for stimulating discussions and guidance in the development of this paper. KN acknowledges support from the Royal Society grant number URF\textbackslash R\textbackslash 231006. OL acknowledges the STFC Consolidated Grant ST/R000476/1 and visits to All Souls College and to the Physics Department, Oxford University.

\section*{Data Availability}

All data used in this analysis are produced from publicly available datasets and software packages. The Millennium XXL galaxy lightcone \citep{Smith2022} can be downloaded from the Millennium database\footnote{\href{http://icc.dur.ac.uk/data/}{http://icc.dur.ac.uk/data/}} and random walk L\'{e}vy-Flight distributions can be reproduced using the publicly available python package \texttt{MiSTree} \citep{NaidooMistree2019}.



\bibliographystyle{rasti}
\bibliography{bibfile} 








\bsp	
\label{lastpage}
\end{document}